\documentclass[conference]{IEEEtran}
\IEEEoverridecommandlockouts
\usepackage{cite}
\usepackage{url}
\usepackage{pgfplots}
\usepackage{tikz}
\usepackage{algorithm}
\usepackage{algpseudocode}

\usepackage{multirow}
\usepackage{amsmath,amssymb,amsfonts}
\usepackage{graphicx}
\usepackage{textcomp}
\usepackage{xcolor}
\def\BibTeX{{\rm B\kern-.05em{\sc i\kern-.025em b}\kern-.08em
    T\kern-.1667em\lower.7ex\hbox{E}\kern-.125emX}}
\begin{document}

\title{Confidential and Protected Disease Classifier using Fully Homomorphic Encryption\\

}
\author{
\IEEEauthorblockN{Aditya Malik, Nalini Ratha, Bharat Yalavarthi, Tilak Sharma, Arjun Kaushik}
\IEEEauthorblockA{
\textit{University at Buffalo, The State University of New York, USA} \\
\{amalik28, nratha, byalavar, tilaksha, kaushik3\}@buffalo.edu}
\and

\and
\IEEEauthorblockN{Charanjit Jutla}
\IEEEauthorblockA{\textit{IBM Research, USA} \\
csjutla@us.ibm.com}
\and
}

\maketitle

\begin{abstract}

With the rapid surge in the prevalence of Large Language Models (LLMs), individuals are increasingly turning to conversational AI for initial insights across various domains, including health-related inquiries such as disease diagnosis. Many users seek potential causes on platforms like ChatGPT or Bard before consulting a medical professional for their ailment. These platforms offer valuable benefits by streamlining the diagnosis process, alleviating the significant workload of healthcare practitioners, and saving users both time and money by avoiding unnecessary doctor visits. However, Despite the convenience of such platforms, sharing personal medical data online poses risks, including the presence of malicious platforms or potential eavesdropping by attackers. To address privacy concerns, we propose a novel framework combining FHE and Deep Learning for a secure and private diagnosis system. Operating on a question-and-answer-based model akin to an interaction with a medical practitioner, this end-to-end secure system employs Fully Homomorphic Encryption (FHE) to handle encrypted input data. Given FHE's computational constraints, we adapt deep neural networks and activation functions to the encryted domain. Further, we also propose a faster algorithm to compute summation of ciphertext elements. Through rigorous experiments, we demonstrate the efficacy of our approach. The proposed framework achieves strict security and privacy with minimal loss in performance.

\end{abstract}


\section{Introduction}

The landscape of healthcare diagnosis has undergone a profound transformation with the integration of advanced technology and data-driven approaches. Traditionally, diagnosis relied on symptom assessment, medical history, and tests, but the advent of intelligent diagnosis systems, powered by Deep Learning, has revolutionized the process. Many individuals resort to seek potential causes using Large Language Models (LLMs) like ChatGPT or Bard before consulting a medical professional for their ailment. Rather than replacing healthcare professionals, these systems serve as invaluable tools, augmenting their expertise and enabling more accurate and efficient identification of medical conditions. This collaborative approach is particularly crucial during challenging circumstances such as pandemics \cite{wosik2020telehealth}, where resources may be limited, as it empowers healthcare professionals to deliver timely and effective care to patients by enhancing diagnostic accuracy and efficiency.



\begin{figure}[htbp]
\centerline{\includegraphics[width=0.48\textwidth]{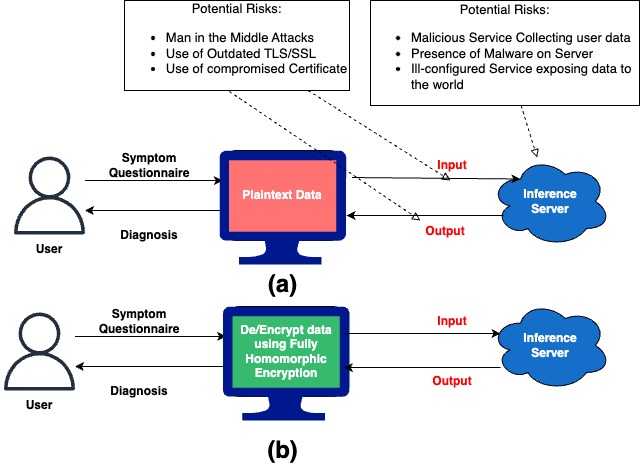}}
\caption{(a) Threat in the current system; (b) Proposed Fully Homomorphic Enryption-enabled Secure Disease Predictor.}
\label{Threats}
\end{figure}

Telehealth \cite{wosik2020telehealth} is an intelligent diagnostic system aimed at reducing a doctor's workload, and helping patients find the most suitable healthcare provider. Telehealth circumvents the need to consult a general physicians in person by providing patients with a system that suggests potential illness based on symptoms. This empowers patients to make informed appointments and navigate their healthcare journey more effectively. But, Telehealth fails to provide a secure system.

As this exchange of information involves user-specific sensitive healthcare data, any diagnosis system must prioritize patient privacy. To this end, various architectures have been developed using encryption to enhance privacy. One such example is the Intel Device Attestation \cite{intel_device}. But, these systems are not end-to-end secure and are susceptible to attacks such as eavesdropping / man-in-the-middle. HTTPS offers a small degree of security when these systems are deployed as cloud servers. But, these cloud-deployed systems are still susceptible to man-in-the-middle attacks or use of outdated TLS / SSL or compromised certificates as shown in Fig. \ref{Threats}(a).




An effective approach to mitigate these risks is to perform computations on encrypted inputs as shown in Fig. \ref{Threats}(b). Fully Homomorphic Encryption (FHE) enables us to perform computations on encrypted inputs and derive results without decryption. FHE ensures that the encryption key remains solely in the hands of the user, guaranteeing exclusive access to the results of the system. This enhances privacy and security, and thereby, maintains the confidentiality of sensitive medical information throughout the diagnostic process.


\textbf{Contributions.} In this paper, we present the development of a secure disease classifier designed to process encrypted patient data and deliver a diagnosis to the patient. While FHE allows computations on encrypted data, certain computational operators are yet to be implemented. We adapt the full connected neural network to the encrypted domain as matrix multiplication. Additionally, we present a more efficient technique for summing ciphertext elements. Further, we approximate activation functions as comparators or polynomial approximations. 


The rest of the paper is organized as follows - Section II touches upon the FHE basics and Section III dives into the related works. Section IV provides a quick overview of our dataset. Section V highlights how each component of the Deep Learning model is uniquely adapted for encrypted data handling. In sections VI and VII, we provide a detailed analysis of our experiments and draw conclusions based on our findings.

\section{FHE Basics}

Homomorphic Encryption is a cryptographic technique that allows computations to be performed on encrypted data without decrypting it first. In other words, it enables computations on data while it remains in its encrypted form, maintaining privacy and confidentiality. There are various types of Homomorphic Encryption schemes, each with its own properties and use cases. Operations supported in Homomorphic Encryption are (Note - E(x) corresponds to encryption of x):

\begin{equation}
E(a)+E(b)=E(a+b)\label{he-add}
\end{equation}
\begin{equation}
E(a)*E(b)=E(a*b)\label{he-mult}
\end{equation}


The most common types include Partially Homomorphic Encryption (PHE) which supports the computation of only one type of operation on encrypted data, either addition or multiplication, Somewhat Homomorphic Encryption (SHE) extends the capabilities of PHE by allowing both addition and multiplication operations on encrypted data. However, there is a limit to the number of operations that can be performed before decryption is required, and Fully Homomorphic Encryption (FHE) allows arbitrary computations to be performed on encrypted data without the need for decryption at any stage. While FHE is highly versatile, it comes with greater computational overhead and complexity.

Fully Homomorphic Encryption (FHE) schemes differ in how they handle computations, and these distinctions often arise from the underlying mathematical structures and optimizations employed. The Brakerski-Vaikuntanathan-Vaikuntanathan (BFV) \cite{brakerski2011efficient} supports integers, while the Cheon-Kim-Kim-Song (CKKS) \cite{cheon2017homomorphic} scheme extends its support to floating-point decimals. The choice of an FHE scheme often depends on the nature of computations required by the application and the balance between security and computational performance sought by the user.

Single Instruction Multiple Data (SIMD) architectures can be leveraged to enhance the efficiency of FHE operations by concurrently processing multiple encrypted data elements. This combination is particularly relevant in scenarios where privacy is a paramount concern, and computations on encrypted data need to be performed with optimal throughput and efficiency. The collaboration of FHE and SIMD contributes to the advancement of secure and efficient computations in privacy-preserving applications.

Our work utilizes HEAAN \cite{heaan}, based on the CKKS scheme, to enable secure computation on encrypted data. In HEAAN, ${log_2 p}$ and ${log_2 q}$ are parameters defining the bit length of the plaintext modulus ${p}$ and the ciphertext modulus ${q}$ respectively. The ${log_2 p}$ value determines the precision of computations, with a larger ${log_2 p}$ providing higher precision. On the other hand, the ${log_2 q}$ value influences the security level, whereas a larger ${log_2 q}$ enhances security. Balancing these parameters is crucial, as it involves optimizing precision and security to meet the specific requirements of the application while managing computational efficiency. 

Although FHE allows computations on encrypted data, several trivial computational operators are yet to be implemented. In this paper, we address this gap by developing FHE-compatible operators, using HEAAN for implementation, specifically for the task of disease diagnosis.

\section{Related Work}

\subsection{Healthcare}

\begin{figure*}
\centerline{\includegraphics[width=0.85\textwidth]{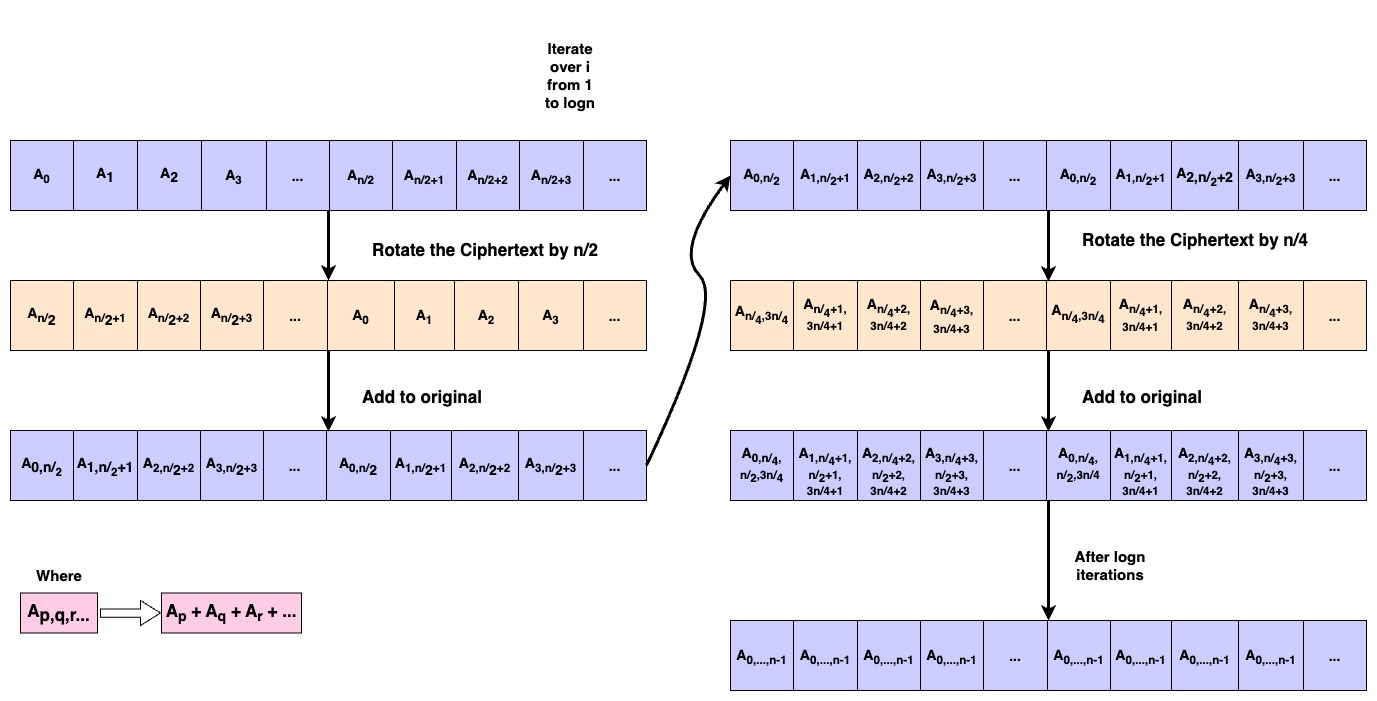}}
\caption{An illustration of the summation of ciphertext elements using \textit{Rotate and Add algorithm}.}
\label{Fully-Connected-Layer}    
\end{figure*}

The challenge of disease diagnosis is critical for healthcare professionals, as inaccuracies could lead to significant complications. Various studies have proposed applications centered around rule-based systems in the clinical domain. For instance, in \cite{imanov2019rule}, the authors introduce a system that diagnoses diabetes using fuzzy logic and a set of expert-defined rules known as VP-Expert. This system aims to enhance accessibility to diabetes diagnosis, particularly in resource-limited settings, facilitating streamlined diagnosis and efficient healthcare delivery \cite{chatgpt}.

In a different study \cite{jariwala2023respiratory}, the authors employ a fuzzy rule-based system to evaluate the risk of respiratory diseases among waste workers. The validation of this system involves the use of Pulmonary Function Tests (PFT) \cite{bhakta2022addressing}, and the results demonstrate the system's effectiveness in predicting levels of respiratory risk. These rule-based approaches contribute to more accessible and streamlined disease diagnosis in specific healthcare contexts.

Beyond the aforementioned approaches, numerous studies delve into disease prediction by employing not only Neural Networks but also various other techniques. The "Intelligent Heart Disease Prediction System" \cite{palaniappan2008intelligent} investigates the application of data mining techniques for disease diagnosis. Another study \cite{afrash2022machine} explores the use of a Clinical Decision Support System (CDSS) to automate COVID-19 diagnosis. Additionally, there are research efforts leveraging model ensemble and reinforcement learning, as exemplified by works such as \cite{cunningham2000stability} \cite{zhong2023hierarchical} \cite{west2005ensemble}. These diverse methodologies contribute to the exploration and advancement of disease prediction models.

Nevertheless, the stringent security guidelines mandated by the Health Insurance Portability and Accountability Act (HIPAA) \cite{hipaa} and the Health Information Technology for Economic and Clinical Health (HITECH) Act \cite{hitechact} impose rigorous standards for the protection and security of healthcare data. To enhance security measures, we investigated the application of Fully Homomorphic Encryption.

\subsection{Fully Homomorphic Encryption in Healthcare}

Homomorphic Encryption for medical diagnosis: secure and privacy-preserving Homomorphic Encryption (HE) has emerged as a promising tool for revolutionizing medical diagnosis by enabling secure and privacy-preserving analysis of sensitive patient data. 

A noteworthy example of Homomorphic Encryption (HE)-based disease classifiers is highlighted in a paper \cite{shaikh2021sensitivity}, which delves into the use of HE for secure analysis of electrocardiogram (ECG) data, paving the way for privacy-preserving heart disease diagnosis. Another area of significant research focuses on secure cancer prediction \cite{sarkar2022scalable}, potentially expanding into specific use cases targeting particular body parts. An exemplar of such a specialized classifier is presented in the work by Son et al. (2021) \cite{son2021privacy}, where they devised a system for the encrypted classification of breast cancer. Similarly, another instance is found in the work by the authors of a paper \cite{adhikary2023secret}, introducing an innovative HE-based Lung Cancer Diagnosis and classifier. This system initially performs textual extraction of CT scans, subsequently employing Deep Learning techniques for classification.

As the generation of Electronic Health Records (EHRs) continues to rise, extensive research is underway to leverage them securely for disease diagnosis. An intriguing illustration involves encrypting EHRs and uploading them to the cloud, where encrypted records are employed to inform users of their likelihood of experiencing cardiovascular disease \cite{bos2014private} \cite{boomija2022secure}. Similarly, Tuong et al. (2022) \cite{nguyen-van2022homomorphic} introduced a Homomorphic Encryption-powered classifier designed to identify mental health conditions based on encrypted phone usage patterns, showcasing the considerable potential for secure mental health analysis. Enhancing the security and accessibility of EHRs involves utilizing blockchain to make the data readily available, coupled with Homomorphic Encryption to ensure the security and privacy of the data \cite{li2022ehrchain}.

\section{Dataset}

"Hierarchical Reinforcement Learning for Automatic Disease Diagnosis" \cite{zhong2023hierarchical} employs a dataset sourced from the SymCat symptom-disease database. Organized into 9 departmental groups, each containing 10 diseases based on the International Classification of Diseases, the dataset encompasses a total of 90 diseases for the classifier to operate on.

\begin{figure}[!ht]
\centerline{\includegraphics[width=60mm]{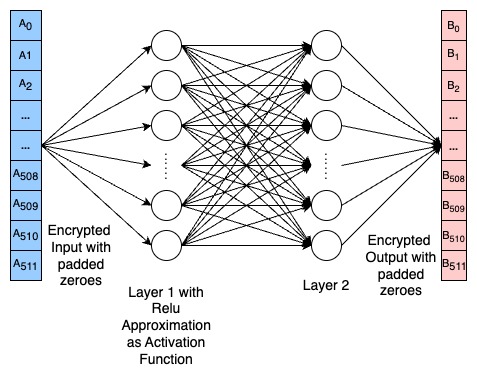}}
\caption{High-level architecture diagram of Fully Homomorphic Encryption-enabled Secure Disease Predictor containing fully connected layer and activation functions.}
\label{Network-Architecture}    
\end{figure}

Following the CDC database \cite{zhong2023hierarchical}, each disease is linked to a distinct set of symptoms. The disease classifier processes a set of explicit and implicit symptoms encrypted as a ciphertext. Upon computation, the user decrypts the received ciphertext to identify the associated disease. In our experiments, participants respond to a survey featuring 266 symptoms. The system predicts one of the 90 possible diseases based on their input. A subset of user symptoms is detailed in Table \ref{symptom_table}.

\begin{table}[htbp]
\caption{Symptom Table (Input).}
\begin{center}
\scalebox{0.9}{
\begin{tabular}{|c|c|}
\hline
\textbf{Symptom} & \textbf{User response} \\
\hline
...& ... \\
\hline
Skin Rash& No \\
\hline
Neck Pain& No \\
\hline
Anxiety and Nervousness & Yes \\
\hline
Depression or Psychotic Symptoms& No \\
\hline
Abnormal Involuntary Movements & No \\
\hline
Eye Redness & No \\
\hline
...& ... \\
\hline
\end{tabular}
}
\label{symptom_table}
\end{center}
\end{table}

\section{Proposed Approach}


FHE ensures the security and privacy of patient data by entrusting the secret key exclusively to the encrypting party, in this case, the patient. Subsequent operations are conducted on the encrypted data until the results are provided back to the patient for decryption, maintaining confidentiality throughout the process. HEAAN \cite{heaan} only allows encryption of data in sizes of powers of 2. We apply zero-padding to both the input and weight matrices, extending them to the size of 512 (a power slightly greater than the input size of 266). To achieve secure disease classification, the following components have been adapted to FHE - (i) Fully connected layer; (ii) ReLU. 


\subsection{Fully connected layer}

 In the plaintext domain, deep neural networks are seamlessly imported, and third-party libraries manage underlying processes. On the other hand, FHE requires manual handling. Fully connected layer is adapted as matrix multiplication to the encrypted domain.

Matrix multiplication involves multiplication and summation. But, summation of ciphertext elements is not straightforward as FHE does not allow access to individual elements. To address this, we implemented algorithm \ref{alg:rotAdd}, closely similar to \cite{hong2022secure}, streamlining the summation within a node of a fully connected layer through rotation and addition, leveraging the principles of CKKS scheme \cite{cheon2017homomorphic}.

\begin{algorithm}
\caption{Rotate and Add}\label{alg:rotAdd}
\begin{algorithmic}[1]
    \Function {RotAdd}{$c$,$size$}
        \State $k$ $\gets$ $\log_2 size$
        \For{$i$ $\gets$ $k-1$ to $0$}
            \State $c_{rot}$ $\gets$ $Rot(c,2^i)$
            \State $c$ $\gets$ $Add(c,c_{rot})$
            \Comment{Ciphertext rotation and addition supported by FHE}
        \EndFor
        \State \Return $c$
    \EndFunction
\end{algorithmic}
\end{algorithm}

To execute the addition after matrix multiplication with the weights, we start by duplicating our original ciphertext of size {$n$}. Then, we rotate the duplicate by {$\frac{n}{2}$ before adding it to the original ciphertext. This results in the first {$\frac{n}{2}$}  and last {$\frac{n}{2}$}  being symmetrical, as depicted in Fig. \ref{Fully-Connected-Layer}, with {$i$} equal to 1. However, achieving the sum of all elements entails repeating this process {$log_2 n$} times, as outlined in algorithm \ref{alg:rotAdd}. Following the iterative rotation and addition of the ciphertext, the total sum is computed across all positions in the ciphertext. 

The multiplication of ciphertext and the rotation and addition method suggested in algorithm \ref{alg:rotAdd} 

Given that the single output from each node in the linear layer is employed to construct the ciphertext for the subsequent layers, a node positional multiplication is indispensable (i.e., multiplying with a zero vector where the value is 1 only at the node index) for the output. This computation is performed for each node, and the outcomes are summed up using the addition property available in the CKKS scheme \cite{cheon2017homomorphic}.

\subsection{LeakyReLU}

Activation functions are essential in deep neural networks to introduce non-linearity. In the encrypted domain, non-linear functions cannot be adapted directly and have to be approximated. We train a polynomial regression model to derive a 8-degree polynomial approximation of LeakyReLU as shown in algorithm \ref{alg:LeakyReLU}. However, the model performance dropped significantly, forcing us to turn to ReLU. We retrain our disease classifier model in the plaintext domain with an approximation of ReLU \cite{sharma2016fully} and adopt the approximate ReLU function. The ReLU approximation is implemented through algorithm \ref{alg:ReluComp}, wherein the comparison between variables 'a' and 'b' results in 0 if 'a' is greater, 1 if 'b' is greater, and 0.5 otherwise.

\begin{algorithm}
\caption{LeakyReLU Approximation
}\label{alg:LeakyReLU}
\begin{algorithmic}[1]
\Function{LeakyReLU Approximation}{$a$}
\State $value$ $\gets$ -$2.42$$x^8$ +$0.28$$x^7$ +$5.32$$x^6$ -$0.51$$x^5$ -$4.10$$x^4$ +$0.28$$x^3$ +$1.60$$x^2$ +$0.51$$x$ +$0.02$ 
\State \Return $value$
\EndFunction
\end{algorithmic}
\end{algorithm}

\begin{algorithm}
\caption{ReluComp \cite{sharma2016fully} }\label{alg:ReluComp}
\begin{algorithmic}[1]
\Function{compB}{$a$,$b$,$n$,$d_g$,$d_f$}
\State $x$ $\gets$ $a$ - $b$
\For{$i$ $\gets$ $1$ to $d_g$}
    \State $x$ $\gets$ $g(n,x)$
\EndFor
\For{$i$ $\gets$ $1$ to $d_f$}
    \State $x$ $\gets$ $f(n,x)$
\EndFor
\State \Return $\frac{x+1}{2}$

\EndFunction
\vspace{.1cm}
\State $g_1(x)$ = -$\frac{1359}{2^{10}}$$x^3$ + $\frac{2126}{2^{10}}$$x$
\vspace{.1cm}
\State $g_2(x)$ = $\frac{3796}{2^{10}}$$x^5$ -  $\frac{6108}{2^{10}}$$x^3$ + $\frac{3334}{2^{10}}$$x$
\vspace{.1cm}
\State $g_3(x)$ = -$\frac{12860}{2^{10}}$$x^7$ + $\frac{25614}{2^{10}}$$x^5$ - $\frac{16577}{2^{10}}$$x^3$ + $\frac{4589}{2^{10}}$$x$
\vspace{.1cm}
\State $g_4(x)$ = $\frac{46623}{2^{10}}$$x^9$ - $\frac{113492}{2^{10}}$$x^7$ + $\frac{97015}{2^{10}}$$x^5$ - $\frac{34974}{2^{10}}$$x^3$ + $\frac{5850}{2^{10}}$$x$
\vspace{.05cm}
\State $f_1(x)$ = -$\frac{1}{2}$$x^3$ + $\frac{3}{2}$$x$
\vspace{.1cm}
\State $f_2(x)$ = $\frac{3}{8}$$x^5$ -  $\frac{10}{8}$$x^3$ + $\frac{15}{8}$$x$
\vspace{.1cm}
\State $f_3(x)$ = -$\frac{5}{16}$$x^7$ + $\frac{21}{16}$$x^5$ - $\frac{35}{16}$$x^3$ + $\frac{35}{16}$$x$
\vspace{.1cm}
\State $f_4(x)$ = $\frac{35}{128}$$x^9$ - $\frac{180}{128}$$x^7$ + $\frac{378}{128}$$x^5$ - $\frac{420}{128}$$x^3$ + $\frac{315}{128}$$x$
\vspace{.1cm}
\end{algorithmic}
\end{algorithm}

\section{Results}

\begin{figure}[!ht]
\centerline{\includegraphics[width=70mm]{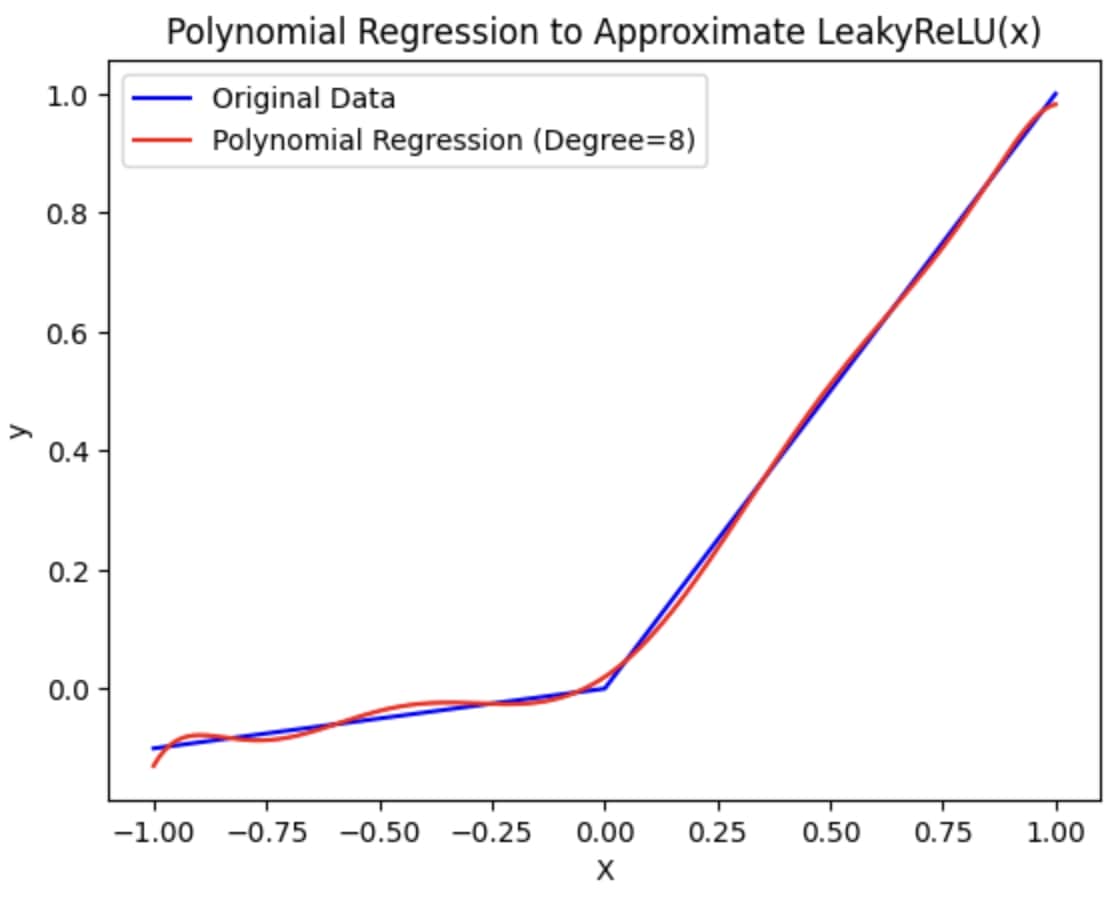}}
\caption{A graph of exact Leaky ReLU (Original Data) and 8-degree polynomial approximation of Leaky ReLU.}
\label{Leaky-Relu}    
\end{figure}

\begin{table*}[!ht]
\caption{Model accuracy in different domains when subjected to different activation functions.}
\begin{center}
\scalebox{1.1}{
\begin{tabular}{|c|c|c|c|}
\hline
\textbf{Activation Function} & 
\textbf{Domain} & 
\textbf{Compared Against} & \textbf{Accuracy} \\
\hline
\begin{tabular}{@{}c@{}}Leaky ReLU\end{tabular}&Plaintext&Test data& 50.40\% \\
\hline
\begin{tabular}{@{}c@{}}Leaky ReLU approximation\end{tabular}&Plaintext&Test data& 39.32\% \\
\hline
\begin{tabular}{@{}c@{}}ReLU approximation \end{tabular}&Plaintext&Test data& 45.02\% \\
\hline
 \begin{tabular}{@{}c@{}}ReLU approximation\end{tabular}& Encrypted&Test data& 45.02\% \\
\hline
\begin{tabular}{@{}c@{}}ReLU approximation \end{tabular}& Encrypted &Plaintext network output on test data & 95.00\% \\
\hline
\end{tabular}
}
\label{result_classifier_accuracy}
\end{center}
\end{table*}

\begin{table*}[!ht]
\caption{Mean absolute error (MAE) of data when subjected to different activation functions during training and inferencing (in the encrypted domain). }
\begin{center}
\scalebox{1.1}{
\begin{tabular}{|c|c|c|}
\hline

\textbf{Train activation function} & \textbf{Inference activation function} & \textbf{MAE} \\
\hline
\begin{tabular}{@{}c@{}}Leaky ReLU \end{tabular}&Leaky ReLU approximation & 0.110 \\
\hline
\begin{tabular}{@{}c@{}}Leaky ReLU  approximation\end{tabular}& Leaky ReLU approximation & 0.090 \\
\hline
\begin{tabular}{@{}c@{}}ReLU  \end{tabular}& ReLU approximation & 0.080\\
\hline
\begin{tabular}{@{}c@{}} ReLU approximation  \end{tabular}&ReLU approximation & 0.017  \\
\hline
\end{tabular}
}
\label{result_activation_accuracy}
\end{center}
\end{table*}

\subsection{Fully connected layer}


We compare algorithm \ref{alg:rotAdd} with the existing method for computing the summation of ciphertext elements, using Discrete Fourier Transform (DFT) \cite{han2019improved}. This approach involves extracting the value at the 0th index, representing the sum of all frequencies (i.e., the total sum), and then adjusting the value to its appropriate position.

\begin{figure}[!ht]
  \centering
  \begin{tikzpicture}
    \begin{axis}[
      width=0.45\textwidth,
      height=0.35\textwidth,
      xlabel={Ciphertext length ($2^x$)},
      ylabel={Percentage},
      font=\bfseries,
      legend pos=north west,
      legend style ={font=\small},
      xmin=3, xmax=12,
      ymin=5, ymax=3000,
      xmode=log,
      xtick={3, 4, 5, 6, 7, 8, 9,10, 11, 12},
      xticklabels={3, 4, 5, 6, 7, 8, 9,10, 11, 12},
      grid=both,
      grid style={line width=.1pt, draw=gray!10},
      major grid style={line width=.2pt,draw=gray!50},
      minor tick num=4,
    ]
    
es
    \addplot[mark=*, blue, line width = 0.5pt] table[x=Ciphertext, y=Error] {
      Ciphertext Error
        3 54.0085
        4 5.56393
        5 280.867
        6 550.884
        7 360.715
        8 431.906
        9 10.2816
        12 2655.09
    };
    \addlegendentry{Relative error percentage}
    
    \addplot[mark=*, red, line width = 0.5pt] table[x=Ciphertext, y=Speedup] {
      Ciphertext Speedup
        3 122.245
        4 138.9
        5 73.406
        6 140.216
        7 203.902
        8 185.171
        9 109.212
        12 99.2016
    };
    \addlegendentry{Relative speedup percentage}

    \end{axis}
  \end{tikzpicture}
  \caption{Summation of ciphertext elements can be performed using two methods - (i) Discrete Fourier Transform (DFT); (ii) Rotate and Add (RA). This plot depicts relative error percentage and relative speedup percentage of these methods for various ciphertext lengths in summing ciphertext elements. Relative error percentage = $\frac{DFT - RA}{RA} * 100$. Relative speedup percentage = $\frac{T_{DFT} - T_{RA}}{T_{RA}} * 100$, where $T$ denotes time taken using the respective method. }
  \label{error-speedup-linear}
\end{figure}
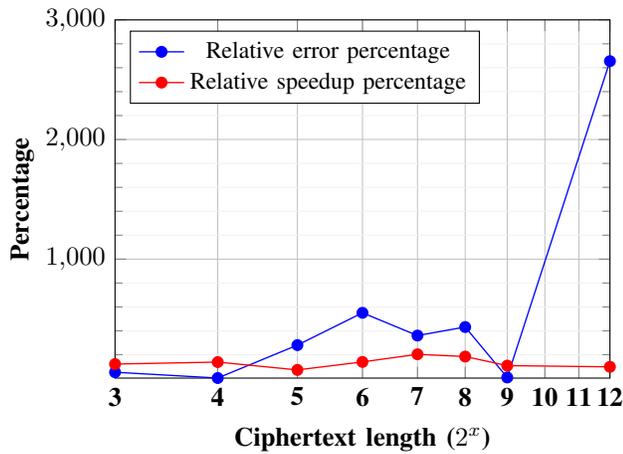


Based on our experiments as shown in Fig. \ref{error-speedup-linear}, algorithm \ref{alg:rotAdd} consistently demonstrates lower errors compared to DFT. It also showcases an \textbf{average decrease in relative error of 543.67\%}. Importantly, the observed errors consistently remain within the range of ${\varepsilon ^{-07}}$.

Both algorithms are further evaluated on their execution times. In the DFT algorithm, the execution time depends on ${r\log_2 n}$, where the optimal value for $r$ is experimentally determined. From Fig. \ref{error-speedup-linear}, we can say that algorithm \ref{alg:rotAdd} outperforms the DFT algorithm across all ciphertext sizes, achieving an impressive \textbf{average relative speedup of 134.03\%}.

\subsection{Leaky ReLU Approximation}

As illustrated in Fig. \ref{Leaky-Relu}, our approximation closely follows the curve of Leaky ReLU. However, in the vicinity of zero (where a significant portion of the input to the activation function is concentrated), the approximation deviates from the curve, leading to increased errors by inverting the signs of negatives to positives, consequently yielding incorrect predictions. Table \ref{result_activation_accuracy} shows a comparative analysis of different activation functions and errors in FHE.

\subsection{Disease Classifier}

Table \ref{result_classifier_accuracy} provides an overview of our experiments and results. We observe a significant drop in model accuracy with approximated Leaky ReLU activation function. ReLU approximation works equally well in both plaintext domain and encrypted domain as evidenced in Table \ref{result_classifier_accuracy}.

\section{Conclusion}

In this paper, we propose to use a combination of Fully Homomorphic Encryption (FHE) and Deep Learning to develop a private and secure diagnosis system. We show that the introduction of FHE does not harm the model's accuracy. Furthermore, the classification performance in the encrypted domain aligns closely with the performance in the unencrypted domain, achieving an accuracy of 95\%. Since FHE cannot execute all computational operators, we detail the adaptation of each component to the encrypted domain. Additionally, we show that our proposed algorithm \ref{alg:rotAdd} excels in precision and speed. It also requires a lower ${\log_2 q}$ value (with respect to ${\log_2 p}$), facilitating efficient computations, and accelerating the entire process. We believe our work can be improved by employing parallelization techniques. There is also tremendous potential to enhance this framework by expanding the scope of diseases and transforming it into more of a dialogue-based system rather than a questionnaire-based one, as it currently stands.

\bibliographystyle{unsrt}
\bibliography{secure}

\vspace{12pt}
\end{document}